\documentclass[superscriptaddress,groupedaddress,nofootnoteinbib,12pt]{article}  

\usepackage{amsmath}
\usepackage{amsfonts}
\usepackage{authblk}
\usepackage{bbm}
\usepackage{color}
\usepackage[dvipsnames]{xcolor}
\usepackage{graphicx,graphics}
\usepackage{epstopdf}
\usepackage{caption}
\usepackage{subcaption}
\usepackage{float} 
\usepackage{pifont}
\usepackage{titlesec}
\usepackage{etoolbox}
\usepackage{indentfirst}
\usepackage{jcappub}
\usepackage{epstopdf}

\def\bb{\begin{equation}}
\def\ee{\end{equation}}

\def\5{\hspace*{5mm}}
\def\2{{\scriptstyle\frac12}}

\begin{document}

\title{Unitarity and Different Phases of the Extended New Massive Gravity}

\author{Giorgi Tukhashvili}
\affiliation{Center for Cosmology and Particle Physics, Department of Physics, \\
New York University, 726 Broadway, New York, NY, 10003, USA}

\emailAdd{gt783@nyu.edu}


\abstract{We study the New Massive Gravity extended by the curvature cubed invariant and the Cosmological Constant using the tree-level exchange amplitudes on a maximally symmetric space-time. We identify the parameter space, consisting of the ratio between C.C. and graviton mass squared and the relative coupling strength between curvature squared and cubed invariants, for which the massive spin-2 is a ghost. Different phases of this model are also discussed.}

\maketitle


\section{Introduction and Summary}


Higher curvature modifications of General Relativity (GR) are important in order to understand the quantum nature of gravity. It has been long known that GR in $3+1$ dimensions is perturbatively non-renormalizable \cite{tHooft:1974toh}. Adding terms that are quadratic in curvature solves this problem at the expense of introducing a massive scalar and a massive spin-2 ghost \cite{Stelle:1976gc}.

In $2+1$ dimensional space-time GR does not suffer the non-renormalizability problem, simply because it does not propagate any degrees of freedom (d.o.f.) \cite{Deser:1983tn}. The massive spin-2 generated through higher curvature modifications can be healthy if one chooses the massless mode to be ghost. The wrong sign kinetic term of the massless spin-2 does not lead to any problems (not considering non-positive central charges of the dual CFTs \cite{Bergshoeff:2009aq}) since in $3D$ this mode is a pure gauge, but in this case the unitarity is violated through the massive scalar. The conformal mode in GR is a ghost, but a closer look reveals that it's not propagating. Addition of higher curvature terms to the action modifies the dynamics of the conformal mode as well. In general, its equation of motion becomes of quartic order and a massive scalar gets generated. The kinetic terms of massive scalar and the massless spin-2 have the same sign, so if one chooses massless spin-2 to be a ghost, so has to be the massive scalar. In general it seems that one can not build a unitary theory in $2+1$ dimensions by introducing higher curvature invariants.

Ten years ago, Bergshoeff, Hohm \& Townsend found a curvature squared invariant that, linearly, is equivalent to Fierz-Pauli and does not violate unitarity \cite{Bergshoeff:2009hq}. Sinha found its curvature cubed extension \cite{Sinha:2010ai} (see also \cite{Gullu:2010pc}). Within this model, now referred as ``Extended New Massive Gravity (ENMG)", the conformal mode of the metric has the structure of a galileon \cite{Gabadadze:2018hos}, so its equation of motion is of second order. As a consequence, the massive scalar is absent and one is free to choose the massless spin-2 to have the wrong sign kinetic term. The authors of \cite{deRham:2011ca} addressed the non-linear unitarity for the curvature squared term and proved that the model was ghost free to all orders. In \cite{Afshar:2014ffa} it is argued that full ENMG is free of scalar ghosts. There are other works addressing the causality problem for this model \cite{Komada:2019zry,Edelstein:2016nml}. The author of this paper is not familiar with any previous works done that addresses the complete unitarity of ENMG.

While writing this paper, I found out that some of my results coincide with \cite{Afshar:2014ffa}, so I feel it's necessary to point out the differences. First of all, our common results agree with each other. The disjoint part is that they include the square of Cotton tensor, which makes the spectrum richer, but necessarily leading to ghosts and tachyons. My analysis is restricted to (\ref{ENMG}), but it's more detailed. Their analysis relies on the Lagrangian formalism, while I'll be using the equations of motion and classical exchange amplitudes. I'll also identify additional phases missed in \cite{Afshar:2014ffa}.

This paper is structured as follows: In Section \ref{sec_2} I will argue that a model containing any power of Ricci tensor can propagate a maximum of three degrees of freedom in $3D$. Then I will argue that the ``Extended New Massive Gravity" defined by (\ref{ENMG}) propagates only two degrees of freedom. In Section \ref{sec_3} I'll do linear analysis of (\ref{ENMG}) on a maximally symmetric background and show that the model does indeed propagate only one massive spin-2, i.e. 2 d.o.f. In Section \ref{sec_4} different phases of (\ref{ENMG}) are discussed. I will argue that there exists a choice of parameters, within the limits of EFT, for which the massive spin-2 is a ghost.

Conventions: 
The flat metric is mostly negative $\eta_{\mu \nu} = \text{diag} (1,-1,-1)$. Ricci tensor is $R_{\mu \nu} = \partial_\rho \Gamma^\rho_{\mu \nu} + \cdots$. Except Section \ref{sec_4} the Planck mass will be set to one.


\section{Counting the Degrees of Freedom}\label{sec_2}

In what follows, I will call the ``Extended New Massive Gravity" to the model defined by the following action:
\begin{align}\label{ENMG}
\mathcal{S} = \int d^3 x \sqrt{g} \bigg[ & R - \frac{1}{3} \Lambda + \frac{1}{m^2} \left( R_{\mu \nu } R^{\mu \nu } - \frac{3}{8} R^2 \right) \\
\nonumber {} & -\frac{1}{\kappa m^4} \left( R_\lambda^\rho R_\rho^\sigma R^\lambda_\sigma - \frac{9}{8} R R_\lambda^\rho R^\lambda_\rho + \frac{17}{64} R^3 \right) \bigg] .
\end{align}
In the limit when $\Lambda \rightarrow 0$, Minkowski space-time is a solution to the equations of motion. When linearized on this background the spectrum consists of one massless and one massive (with mass $=m$) spin-2 modes. In this section I will argue that this is the case for an arbitrary background.

A general diffeomorphism invariant action involving any degree of curvatures\footnote{Terms involving the derivatives of curvatures are not considered.} can propagate a maximum of three degrees of freedom in $2+1$ dimensions. In order to see this let's decompose the metric in the following way:
\bb
g_{\mu \nu} = \bar{g}_{\mu \nu} + h_{\mu \nu} + \bar{\nabla}_{( \mu} A_{\nu )} + \bar{\nabla}_{\mu} \bar{\nabla}_{\nu} \varphi + e^{2 \pi} \bar{g}_{\mu \nu}.
\ee
Here $\bar{g}_{\mu \nu}$ is arbitrary solution to (\ref{ENMG}) and $\bar{\nabla}$ is covariant derivative with respect to this background. $h_{\mu \nu}$, $A_{\mu} $ and $\varphi$ are tensor, vector and scalar perturbations on this metric (linear response to matter). The last piece, $\exp\left( 2 \pi \right)$, is the conformal mode. Because of the diffeomorphism invariance of the action, neither $A_{\mu} $ nor $\varphi$ can be physical. The equations of motion, in general, involve quartic time derivatives, therefore there are two spin-2 modes and one massive scalar. The massless spin-2 in $2+1$ dimensions is a pure gauge, so we are left with two degrees of freedom associated with massive spin-2 and one d.o.f. coming from the massive conformal scalar.

The combination (\ref{ENMG}) is special in a sense that the conformal mode turns out to be a galileon \cite{Gabadadze:2018hos}. The equations of motion corresponding to the variation of $\pi$ are of second order (w.r.t $\pi$) and the massive scalar disappears from the spectrum, so the model defined by (\ref{ENMG}) propagates only two degrees of freedom. In three dimensions this is the unique combination with non-trivial conformal mode having this property, any higher order $(>3)$ curvature invariant should either have a trivial conformal mode or will propagate three degrees of freedom and will not be unitary.\footnote{In $2+1$ dimensions the square of Cotton tensor $\sqrt{-g} C_{\mu \nu \rho} C^{\mu \nu \rho}$ has a trivial conformal mode. This term is sometimes considered together with (\ref{ENMG}) \cite{Afshar:2014ffa}, I will ignore it since it contains the derivatives of curvature.}

The overall sign of the action in (\ref{ENMG}) guarantees that, when linearized around flat background $(\Lambda=0)$, the kinetic term of the massive spin-2 has the right sign. 
In the following sections I will positively answer the following question: Is there a choice of parameters $(\Lambda/m^2;\kappa)$, within the limits of EFT, on which the kinetic terms of spin-2 modes flip the sign?


\section{Analysis of the Linear Perturbations}\label{sec_3}

\subsection{A Toy Model for Linear Perturbations}

\begin{figure}[H]
	\centering
	\includegraphics[width=0.9\textwidth]{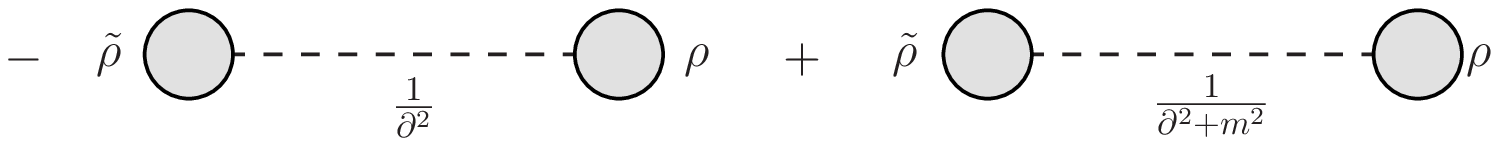}
	\caption{Diagrammatic representation of the exchange amplitude (\ref{toy_amplitude})}\label{diag}
\end{figure}

Understanding the physical meaning of an expression with complex tensorial structure is not an easy task. The analysis gets even more complicated when calculations are done on a curved background. In the next section, while calculating exchange amplitudes, I will rely on the equations of motion and will avoid dealing with Lagrangians. In order to gain some intuition for the conclusions I'm going to make, it is useful to first consider the following toy model of a scalar field:
\bb\label{toy}
\mathcal{L} = \frac12 \phi \partial^2 \phi + \frac{1}{4} a \phi^2 \partial^2 \phi - \frac12 b \left( \partial^2 \phi \right)^2
- \frac12 c \phi \left( \partial^2 \phi \right)^2 .
\ee
Here $a,b,c$ are dimensionfull constants. An arbitrary $\phi = \phi_0 = const$ is a solution of (\ref{toy}). Let's introduce some source $\rho$ and study the linear response on this background $(\phi = \phi_0 + \chi)$:
\bb\label{toy_lin}
\mathcal{L}^{(2)} = \frac12 \left( 1 + a \phi_0 \right) \chi \partial^2 \chi - \frac12 \left( b + c \phi_0 \right)
\left( \partial^2 \chi \right)^2 + \chi \rho .
\ee
Solution to this Lagrangian and the exchange amplitude between two sources are respectively given by:
\bb
\chi = - \frac{1}{1 + a \phi_0} \left[ \frac{1}{\partial^2} - \frac{1}{\partial^2 - \frac{1 + a \phi_0}{b + c \phi_0} } \right] \rho ,
\ee
\bb\label{toy_amplitude}
\mathcal{A} = \int d^3 x \tilde{\rho} \chi = - \frac{1}{1 + a \phi_0} \int d^3 x
\left[ \tilde{\rho} \frac{1}{\partial^2} \rho - \tilde{\rho} \frac{1}{\partial^2 - \frac{1 + a \phi_0}{b + c \phi_0} } \rho \right] .
\ee
Mediators are massless and massive scalars (see Figure \ref{diag}), one of them is necessarily a ghost. When $a \neq 0$ the amplitude has a peculiar overall factor due to the renormalization of the coupling strength on a non-trivial background. The matter of which mode is ghost depends on the sign of this factor. To see this more clearly let's introduce a Lagrange multiplier and rewrite (\ref{toy_lin}) in an equivalent form:
\bb
\mathcal{L}^{(2)} = \frac12 \left( 1 + a \phi_0 \right) \chi \partial^2 \chi - \left( 1 + a \phi_0 \right) \varphi \partial^2 \chi + \frac12 \frac{\left( 1 + a \phi_0 \right)^2}{b + c \phi_0} \varphi^2 + \chi \rho .
\ee
Making a field redefinition, $\chi = \omega + \varphi$, unmixes the two modes from each other:
\bb
\mathcal{L}^{(2)} = \frac12 \left( 1 + a \phi_0 \right) \omega \partial^2 \omega - \frac12 \left( 1 + a \phi_0 \right) \varphi \partial^2 \varphi + \frac12 \frac{\left( 1 + a \phi_0 \right)^2}{b + c \phi_0} \varphi^2 + \omega \rho + \varphi \rho .
\ee
From this expression we see that the factor $\left( 1 + a \phi_0 \right)$ does indeed decide which field should be a ghost. When $\left( 1 + a \phi_0 \right) = 0$, the model is infinitely strongly coupled.

Disclaimer: The model defined by (\ref{toy}) is an EFT with some cutoff. Whether the values of $\phi_0$ for which $\left( 1 + a \phi_0 \right)$ flips sign are within the range of EFT is another story. I will ignore this issue for the present case and will come back to it for ENMG in section \ref{sec_4}.


\subsection{Linearized GR on a Maximally Symmetric Space-Time}

It is instructive to consider the linearized General Relativity on a maximally symmetric background \cite{Porrati:2000cp} before linearizing (\ref{ENMG}). The Einstein-Hilbert action and vacuum equations of motion are:
\bb\label{einstein_hilbert}
\mathcal{S}_{EH} =  - \int d^3 x \sqrt{g} \left( R - \frac{1}{3} \Lambda \right)
\5\5 \rightarrow \5\5
G_{\mu \nu} + \frac{1}{6} \Lambda g_{\mu \nu} = 0 .
\ee
Riemann and Ricci curvatures of a three dimensional maximally symmetric space-time satisfy:
\bb
\bar{R}_{\mu \rho \nu \sigma} = \frac{\bar{R}}{6} \Big( \bar{g}_{\mu \nu} \bar{g}_{\rho \sigma} - \bar{g}_{\mu \sigma } \bar{g}_{\rho \nu} \Big) ,
\5\5\5
\bar{R}_{\mu \nu} = \frac{\bar{R}}{3} \bar{g}_{\mu \nu} .
\ee
From (\ref{einstein_hilbert}) we also have $\bar{R} = \Lambda$. Here $\bar{g}_{\mu \nu}$ is the background metric and the expressions with a bar are defined with respect to it. On the next step let's define the operator that will simplify future expressions:
\bb
\bar{\Box} \equiv \bar{\Delta}_L + \frac{2}{3} \bar{R} .
\ee
$\bar{\Delta}_L$ stands for the Lichnerowicz Laplacian \cite{Lichnerowicz:1964xna} (up to a factor of $-1$) acting on scalar, vector and tensor as:
\bb
\bar{\Delta}_L a = \bar{\nabla}^2 a,
\5
\bar{\Delta}_L b_\mu \equiv \left( \bar{\nabla}^2 - \frac{1}{3} R \right) b_\mu,
\5
\bar{\Delta}_L c_{\mu \nu} \equiv \bar{\nabla}^2 c_{\mu \nu} - \bar{R} c_{\mu \nu} + \frac{\bar{R}}{3} \bar{g}_{\mu \nu} c^\rho_\rho .
\ee
The covariant derivative $\bar{\nabla}_\mu$ and the background metric $\bar{g}_{\mu \nu}$ commute with $\bar{\Box}$. The linear response of spin-2 to matter on this background:
\bb\label{EoM_GR_pert}
\bar{\nabla}_\mu \bar{\nabla}_\rho h_\nu^\rho +
\bar{\nabla}_\nu \bar{\nabla}_\rho h_\mu^\rho - \bar{\nabla}_\mu \bar{\nabla}_\nu h - \bar{\Box} h_{\mu \nu}
= 2 \left( T_{\mu \nu} - \bar{g}_{\mu \nu} T \right).
\ee
Fixing the gauge as:
\bb\label{gauge}
\bar{\nabla}^\nu h_{\mu \nu} = \frac12 \bar{\nabla}_\mu h
\ee
gives the following solution:
\bb
h_{\mu \nu} = - \frac{2}{ \bar{\Box} } \left( T_{\mu \nu} - \bar{g}_{\mu \nu} T \right).
\ee
A covariantly conserved energy momentum tensor is consistent with linear perturbations around a metric that is solution to the Einstein equation in vacuum, meaning
\bb
\mathcal{A} = \int d^3 x \sqrt{\bar{g}} \tilde{T}^{\mu \nu} h_{\mu \nu} =
- \int d^3 x \sqrt{\bar{g}} ~ \tilde{T}^{\mu \nu} \frac{2}{ \bar{\Box} } \left( T_{\mu \nu} - \bar{g}_{\mu \nu} T \right)
\ee
is a gauge invariant amplitude. Even though the pole is shifted, from the tensorial structure it is easy to see that we are dealing with one massless spin-2 particle. Interested reader is invited to check \cite{Porrati:2000cp} to see that $\bar{\Box}$ does indeed have the right properties when acted on transverse and traceless tensors.


\subsection{Linearized ENMG}

In this section we discuss the linearized version of the Extended New Massive Gravity (\ref{ENMG}) on a maximally symmetric space-time. We can write the equation of motion as:
\bb\label{full_EoM}
G_{\mu \nu} + \frac{1}{6} \Lambda g_{\mu \nu} + \frac{1}{m^2} K_{\mu \nu} - \frac{1}{\kappa m^4} Q_{\mu \nu} = 0 ,
\ee
where $K_{\mu \nu}$ and $Q_{\mu \nu}$ are given in Appendix \ref{app_1}. To the lowest order we assume the matter to be absent. Since we are dealing with maximally symmetric space-time, the background satisfies a scalar equation:
\bb\label{background_EoM}
\bar{R}^3 + 8 \kappa m^2 \bar{R}^2 + 192 \kappa m^4 \bar{R} - 192 \kappa m^4 \Lambda = 0 .
\ee
After fixing the gauge (\ref{gauge}), the linear response to matter becomes:
\bb\label{EoM_ENMG}
\gamma \left( \bar{\Box}^2 h_{\mu \nu} - \frac{1}{4} \bar{g}_{\mu \nu} \bar{\Box}^2 h - \frac{1}{4} \bar{\nabla}_\mu \bar{\nabla}_\nu \bar{\Box} h \right) + \beta m^2 \bar{\Box} h_{\mu \nu} - \frac{1}{2} \alpha m^2 \bar{g}_{\mu \nu} \bar{\Box} h = -2 m^2 T_{\mu \nu} ,
\ee
with:
\bb
\alpha = 1 + \frac{\bar{R}}{12 m^2} + \frac{\bar{R}^2}{64 \kappa m^4} ;
\5
\beta = 1 - \frac{\bar{R}}{12 m^2} - \frac{\bar{R}^2}{192 \kappa m^4} ;
\5
\gamma = 1 + \frac{\bar{R}}{8 \kappa m^2} .
\ee
Note that the trace of (\ref{EoM_ENMG}) is of second order as a consequence of the conformal mode being a galileon:
\bb\label{linear_trace_eq}
\alpha \bar{\Box} h = 4 T .
\ee
Since the full theory is gauge invariant, LHS of the equation (\ref{EoM_ENMG}) is consistent with covariantly conserved energy-momentum tensor. Solution to (\ref{EoM_ENMG}) and gauge invariant amplitude are respectively:
\begin{align}
\nonumber h_{\mu \nu } = - \frac{1}{\beta} \Bigg\{ \frac{2}{\bar{\Box}} \left( T_{\mu \nu } - \bar{g}_{\mu \nu} T \right)
& - \frac{2}{\bar{\Box} + \frac{\beta}{\gamma} m^2} \left[ T_{\mu \nu } -
\left( 1 - \frac{\beta}{2 \alpha} \right)
\bar{g}_{\mu \nu} T \right] \\
{} & - \frac{\beta}{\alpha} \bar{\nabla}_\mu \bar{\nabla}_\nu \frac{1}{ \bar{\Box} \left( \bar{\Box} + \frac{\beta}{\gamma} m^2 \right)} T \Bigg\}
\end{align}
\begin{align}
\nonumber \mathcal{A} =  - \int d^3 x \sqrt{\bar{g}} \tilde{T}^{\mu \nu} h_{\mu \nu} = &
\frac{1}{\beta} \int d^3 x \sqrt{\bar{g}} \Bigg\{ \tilde{T}^{\mu \nu} \frac{2}{ \bar{\Box} } \left( T_{\mu \nu} - \bar{g}_{\mu \nu} T \right) \\
{} & - \tilde{T}^{\mu \nu} \frac{2}{\bar{\Box} + \frac{\beta}{\gamma} m^2} \left[ T_{\mu \nu } -
\left( 1 - \frac{\beta}{2 \alpha} \right)
\bar{g}_{\mu \nu} T \right] \Bigg\} .
\end{align}
From the last expression we see that the spectrum consists of massless and massive spin-2 modes. As we saw in the toy model, the matter of which one is ghost and which one is particle depends on the sign of $\beta$.


\section{Different Phases of ENMG}\label{sec_4}

\begin{figure}[t]
	\centering
	\begin{subfigure}[b]{0.49\linewidth}
		\includegraphics[width=\linewidth]{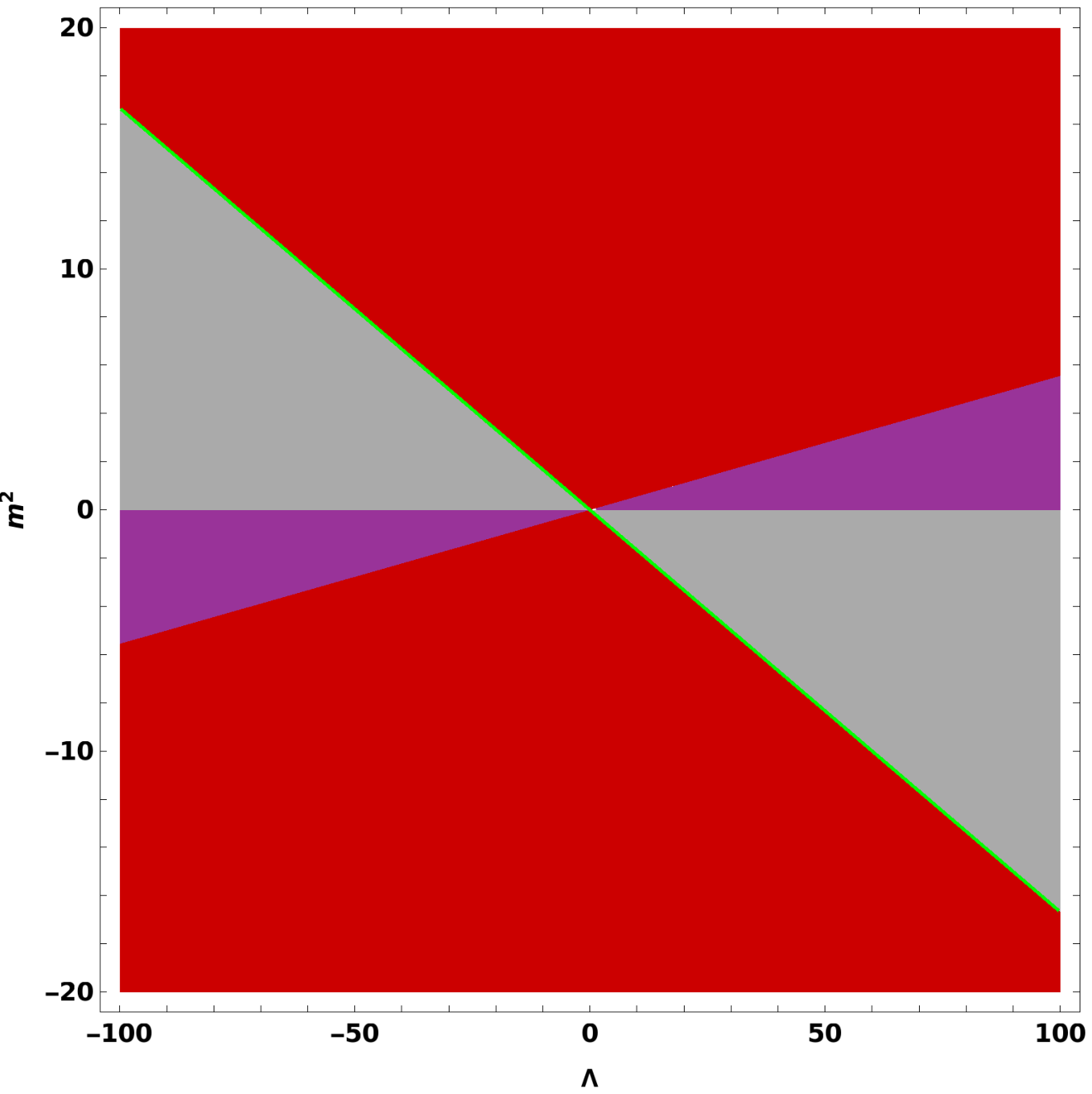}
		\caption{}\label{phase_1}
	\end{subfigure}
	\begin{subfigure}[b]{0.49\linewidth}
		\includegraphics[width=\linewidth]{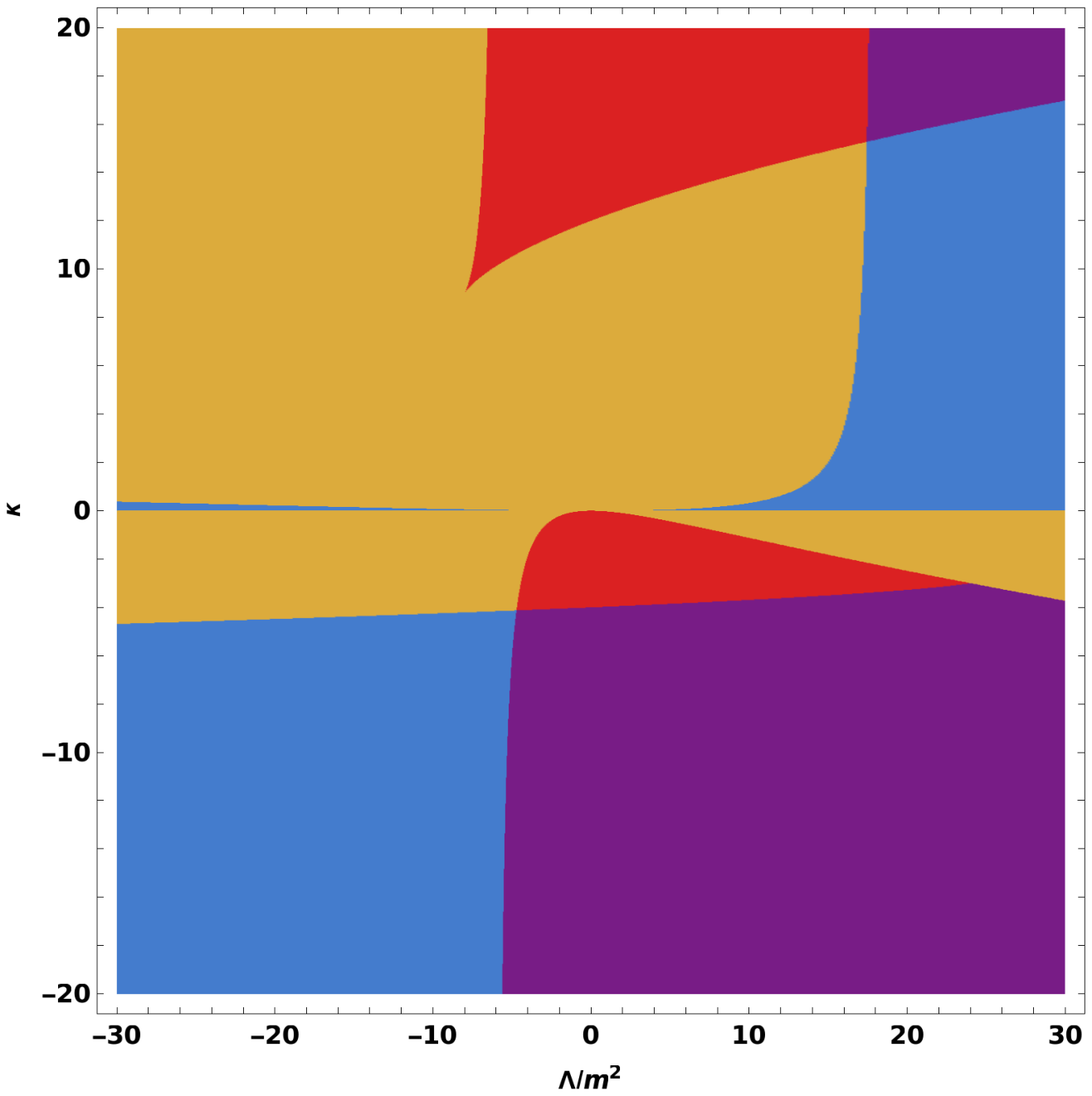}
		\caption{}\label{phase_2}
	\end{subfigure}
	\caption{Left: The New Massive Gravity limit $(\kappa \rightarrow \infty)$. For the red region solutions of (\ref{background_EoM}) are real and corresponding $\beta$s are positive. For the purple region solutions are real and one of the $\beta$s is negative. In the gray region there are no maximally symmetric solutions. The green line corresponds to the enhanced symmetry phase (\ref{enhanced_symmetry}). The units are arbitrary. \\
	Right: Phases of the ENMG. For the red region all solutions of (\ref{background_EoM}) are real and corresponding $\beta$s are positive. For the purple region, all solutions are real and at least one of the $\beta$s is negative. For the orange region one solution of (\ref{background_EoM}) is real and corresponding $\beta$ is positive. For the blue region one solution is real and corresponding $\beta$ is negative. This figure is produced by doing numerical analysis of (\ref{background_EoM}). The discontinuity of thin blue region around $\kappa=0$ is a numerical artifact and it should be connected. }
\end{figure}

Let's discuss how ENMG behaves for the different choice of parameters $m^2,~\Lambda$  and $\kappa$. Since the massless mode does not propagate we can ignore it and concentrate on the massive spin-2. Absence of tachyonic instabilities and dynamical ghosts require $\frac{\beta}{\gamma}m^2 >0$ and $\beta>0$ respectively. First of all let's focus on the case when $\kappa \rightarrow \infty$, i.e. New Massive Gravity amended by the C.C. For this case the above conditions translate to (see Figure \ref{phase_1}):
\bb\label{unit_cond_NMG}
-6 \leq \frac{\Lambda}{m^2} < 18 .
\ee
Lower bound comes by demanding (\ref{background_EoM}) to have real roots. $\Lambda = -6m^2$ is special, at this point $\alpha =0$ and from (\ref{linear_trace_eq}) we see that so does $T$. Linear model acquires scale invariance and diffeomorphism transformations get enhanced by a Weyl piece \cite{Gabadadze:2012xv}:
\bb\label{enhanced_symmetry}
h_{\mu \nu} \rightarrow h_{\mu \nu} + \bar{\nabla}_\mu \xi_\nu +  \bar{\nabla}_\nu \xi_\mu + \pi \bar{g}_{\mu \nu}.
\ee

In the NMG limit $\gamma=1$. When C.C. is absent, $\beta=1$ or $\beta=3$ and around a maximally symmetric background, the model is free of dynamical ghosts and tachyonic instabilities $(m^2>0)$. As we can see from (\ref{unit_cond_NMG}) this is not true for general $\Lambda$. Let's see what changes if we regard (\ref{ENMG}) as an EFT. The lowest cutoff of this EFT is (after recovering the Planck mass) $\Lambda_{5/2} = (\sqrt{M_p} m^2)^{2/5}$ \cite{deRham:2011ca} \footnote{$\Lambda_{5/2}$ is the lowest scale of this model, it appears in the decoupling limit of ``New Massive Gravity" \cite{deRham:2011ca} and defines the cutoff for cubic galileon. 
Even lower cuttoff scale would be $m$, but in this case it makes no sense to talk about the massive spin-2.}.
Recovering the Planck mass does not affect (\ref{unit_cond_NMG}). The values of $\Lambda$ for which $\beta <0$ are within the EFT regime, $\Lambda \ll \Lambda_{5/2}^2$, as long as $10^4 m \ll M_p$. The latter condition seems reasonable and is expected to hold, therefore (\ref{ENMG}) with $\kappa \rightarrow \infty$ can host a massive spin-2 ghost.

The phase diagram for finite $\kappa$ is shown on Figure \ref{phase_2}. Here the red and orange regions are healthy (the kinetic term has the right sign, but it can be tachyonic), for the blue and purple regions we get ghosts. On the boundary between red and purple or orange and blue the model is infinitely strongly coupled. Once again, there exist parameter choices within the EFT for which the massive spin-2 is a ghost, i.e. some parts of the blue and purple regions belong to EFT.

For $\Lambda = -8 \kappa m^2$ there exists a solution of (\ref{background_EoM}) for which $\gamma=0$ and we get an exotic phase. The equation of motion (\ref{EoM_ENMG}) becomes second order and massive spin-2 disappears from the spectrum. From the identity:
\bb
\alpha - \beta = \gamma \frac{\bar{R}}{6 m^2}
\ee
it follows that, when $\gamma=0$, $\alpha = \beta = \left( \kappa+3 \right)/3$, further choosing $\kappa = -3$ complete LHS of (\ref{EoM_ENMG}) vanishes and all the linear dynamics is lost. In Appendix \ref{app_3} I'll show that even though for $\gamma=0$ the massive mode is absent linearly, it reappears non-linearly. Therefore $\gamma=0$ phase is infinitely strongly coupled.


\acknowledgments

I would like to thank Gregory Gabadadze for guidance at all stages of this project and to David Pirtskhalava for discussions and comments. The project was supported by the NYU James Arthur Graduate Fellowship and by the NSF grant PHY-1620039.


\appendix
\section{Appendix: Equations of Motion}\label{app_1}

The variation of the curvature squared and curvature cubed parts of (\ref{EoM_ENMG}) are respectively:
\bb
K_{\mu \nu} = \nabla^2 R_{\mu \nu} - \frac{1}{4} \Big( g_{\mu \nu} \nabla^2 R + \nabla_\mu \nabla_\nu R \Big) - 4 R_{\mu \rho} R^\rho_\nu
+ \frac{9}{4} R R_{\mu \nu} + \frac12 g_{\mu \nu} \Big( 3 R_{\rho \sigma} R^{\rho \sigma} - \frac{13}{8} R^2 \Big) ;
\ee
\begin{align}
\nonumber {} Q_{\mu \nu} = & g_{\mu \nu} \left( \frac{5}{2} R_\lambda^\rho R_\rho^\sigma R^\lambda_\sigma - \frac{51}{16} R R_\lambda^\rho R^\lambda_\rho + \frac{127}{128} R^3 \right)
+ \frac{3}{2} g_{\mu \nu} \nabla^\rho \nabla^\sigma V_{\rho \sigma} + \frac{3}{2} \nabla^2 V_{\mu \nu} \\
{} & - \frac{3}{2} \nabla_\mu \nabla_\rho V^\rho_\nu - \frac{3}{2} \nabla_\nu \nabla_\rho V^\rho_\mu
 - \frac{9}{8} g_{\mu \nu} \nabla^2 \left( R_\lambda^\rho R^\lambda_\rho - \frac{17}{24} R^2 \right) \\
\nonumber {} & + \frac{9}{8} \nabla_\mu \nabla_\nu \left( R_\lambda^\rho R^\lambda_\rho - \frac{17}{24} R^2 \right)
- 6 R^3_{\mu \nu} + 6 R R^2_{\mu \nu} + \frac{15}{8} R_\lambda^\rho R^\lambda_\rho R_{\mu \nu}
- \frac{165}{64} R^2 R_{\mu \nu} ;
\end{align}
where:
\bb
V_{\mu \nu} \equiv R^2_{\mu \nu} - \frac{3}{4} R R_{\mu \nu}
\5\5 \text{and} \5\5
R^{n+1}_{\mu \nu} \equiv R_\mu^{\rho_1} R_{\rho_1}^{\rho_2} \cdots R_{\rho_n \nu } .
\ee
Linearized versions of the different parts of (\ref{full_EoM}):
\bb
G_{\mu \nu} + \frac{1}{6} \Lambda g_{\mu \nu} = \frac12 \Theta_{\mu \nu} - \frac{\bar{R} }{6} \bar{g}_{\mu \nu} h + \frac{1}{6} \left( \Lambda - \bar{R} \right) h_{\mu \nu}
+ \frac{1}{6} \left( \Lambda - \bar{R} \right) \bar{g}_{\mu \nu} ;
\ee
\begin{align}
\nonumber K_{\mu \nu} = & - \frac12 \left[ \Phi_{\mu \nu} + \frac{\bar{R}}{6} \Big( \bar{\nabla}_\mu \bar{\nabla}_\nu h + \bar{g}_{\mu \nu} \bar{\nabla}_\alpha \bar{\nabla}_\beta h^{\alpha \beta} \Big) \right] \\
{} & - \frac{\bar{R}}{24} \Theta_{\mu \nu}
- \frac{\bar{R}^2}{72} \bar{g}_{\mu \nu} h - \frac{\bar{R}^2}{144} h_{\mu \nu} - \frac{\bar{R}^2}{144} \bar{g}_{\mu \nu} ; \\
\nonumber Q_{\mu \nu} = & \frac{\bar{R}}{16} \left[ \Phi_{\mu \nu} + \frac{\bar{R}}{6} \Big( \bar{\nabla}_\mu \bar{\nabla}_\nu h + \bar{g}_{\mu \nu} \bar{\nabla}_\alpha \bar{\nabla}_\beta h^{\alpha \beta} \Big) \right] + \frac{\bar{R}^2}{384} \Theta_{\mu \nu} \\
{} & + \frac{\bar{R}^3}{384} \bar{g}_{\mu \nu} h + \frac{\bar{R}^3}{1152} h_{\mu \nu} + \frac{\bar{R}^3}{1152} \bar{g}_{\mu \nu} ;
\end{align}
with:
\begin{align}
\Theta_{\mu \nu} \equiv & \bar{\nabla}_\mu \bar{\nabla}_\alpha h^\alpha_\nu + \bar{\nabla}_\nu \bar{\nabla}_\alpha h^\alpha_\mu + \bar{g}_{\mu \nu} \bar{\Box} h - \bar{\Box} h_{\mu \nu} - \bar{\nabla}_\mu \bar{\nabla}_\nu h
- \bar{g}_{\mu \nu} \bar{\nabla}_\alpha \bar{\nabla}_\beta h^{\alpha \beta} ; \\
\nonumber \Phi_{\mu \nu} \equiv & \bar{\Box}^2 h_{\mu \nu} - \bar{\Box} \bar{\nabla}_\mu \bar{\nabla}_\alpha h^\alpha_\nu - \bar{\Box} \bar{\nabla}_\nu \bar{\nabla}_\alpha h^\alpha_\mu + \frac12 \bar{\Box} \bar{\nabla}_\mu \bar{\nabla}_\nu h +
\frac12 \bar{g}_{\mu \nu} \bar{\Box} \bar{\nabla}_\alpha \bar{\nabla}_\beta h^{\alpha \beta} \\
{} & + \frac12 \bar{\nabla}_\mu \bar{\nabla}_\nu \bar{\nabla}_\alpha \bar{\nabla}_\beta h^{\alpha \beta}
- \frac12 \bar{g}_{\mu \nu} \bar{\Box}^2 h .
\end{align}
Using these, we can write (\ref{full_EoM}) as:
\bb\label{lin_EoM}
-\frac{\gamma}{m^2} \left[ \Phi_{\mu \nu} + \frac{\bar{R}}{6} \Big( \bar{\nabla}_\mu \bar{\nabla}_\nu h + \bar{g}_{\mu \nu} \bar{\nabla}_\alpha \bar{\nabla}_\beta h^{\alpha \beta} \Big) \right] + \beta \Theta_{\mu \nu} -
\frac{\alpha \bar{R}}{3} \bar{g}_{\mu \nu} h = 2 T_{\mu \nu} .
\ee
Here we used the background equation (\ref{background_EoM}). After fixing the gauge (\ref{gauge}), we recover (\ref{EoM_ENMG}). Setting $\gamma=0$, (\ref{lin_EoM}) reduces to linearized Einstein equation (\ref{EoM_GR_pert}) with rescaled Planck mass.


\section{Appendix: Non-Linear Corrections}\label{app_3}

In order to see why $\gamma=0$ phase is strongly coupled, we can study the next order corrections to the equations of motion. For my purposes, it is enough to study the $\partial^2 h \partial^4 h$ interactions. This sector only appears in $Q_{\mu \nu}$ and its presence is independent of the value of $\gamma$:
\begin{align}
\nonumber Q_{\mu \nu} \supset &  \frac{3}{2} g_{\mu \nu} \nabla^\rho \nabla^\sigma V_{\rho \sigma} + \frac{3}{2} \nabla^2 V_{\mu \nu} - \frac{3}{2} \nabla_\mu \nabla_\rho V^\rho_\nu - \frac{3}{2} \nabla_\nu \nabla_\rho V^\rho_\mu \\
{} & - \frac{9}{8} g_{\mu \nu} \nabla^2 \left( R_\lambda^\rho R^\lambda_\rho - \frac{17}{24} R^2 \right)
 + \frac{9}{8} \nabla_\mu \nabla_\nu \left( R_\lambda^\rho R^\lambda_\rho - \frac{17}{24} R^2 \right) .
\end{align}
Since we are only focusing on the $\partial^2 h \partial^4 h$ interactions, we can set C.C. and the background curvature to zero $\Lambda = \bar{R}=0$ and linearize this expression around the Minkowski space-time. I will work in the transverse and traceless gauge $\partial_\nu h^\nu_\mu =0 = h$. Relevant part of $Q_{\mu \nu}$:
\begin{align}
\nonumber {} Q_{\mu \nu} \supset \frac{3}{8} \bigg( & \partial^4 h_{\mu \rho} \partial^2 h^\rho_\nu
+ \partial^4 h_{\nu \rho} \partial^2 h^\rho_\mu
- \partial_\mu \partial_\rho \partial^2 h_{\nu \lambda} \partial^2 h^{\rho \lambda}
- \partial_\nu \partial_\rho \partial^2 h_{\mu \lambda} \partial^2 h^{\rho \lambda} \\
{} & - \frac{3}{2} \eta_{\mu \nu} \partial^4 h_{\rho \sigma} \partial^2 h^{\rho \sigma}
+ \frac{3}{2} \partial_\mu \partial_\nu \partial^2 h_{\rho \sigma} \partial^2 h^{\rho \sigma} \bigg) .
\end{align}
Equation for the next order corrections contains quartic derivatives. Even though when $\gamma = 0$ the linear equations are of second order and as a consequence the massive mode disappears, it will reappear in the next order perturbations. This shows that $\gamma = 0$ phase is infinitely strongly coupled.



\end{document}